\documentclass[aps,pra,amsfonts,twocolumn,showpacs]{revtex4-1}

\usepackage{amsmath}
\usepackage{amsthm}
\usepackage{bbm}
\usepackage{bm}
\usepackage{braket}
\usepackage{color}
\usepackage{graphicx}
\usepackage{cases}
\usepackage{enumitem}
\usepackage{subfigure}
\usepackage{hyperref}

\renewcommand{\vec}[1]{\bm{#1}}
\DeclareMathOperator{\tr}{tr}
\newenvironment{rcases}{\left.\begin{aligned}}{\end{aligned}\right\rbrace}

\newtheorem{theorem}{Theorem}

\newtheorem*{conjecture*}{Conjecture}
\newtheorem{example}{Example}

\def\E{\cal E}

\def\I{\mathbbm{1}}
\def\wt{\widetilde}

\def\BTB{B^\mathrm{T_B}}
\def\wtBTB{\wt{B}^\mathrm{T_B}}
\def\EW4{\mathrm{EW}_4}

\usepackage{color,soul}
\sethlcolor{yellow}

\begin{document}

\title{Geometric representation of two-qubit entanglement witnesses}
\author{Antony Milne}
\email{antony.milne@gmail.com}
\author{David Jennings}
\author{Terry Rudolph}
\affiliation{Controlled Quantum Dynamics Theory, Department of Physics, Imperial College London, London SW7 2AZ, UK}

\date{\today}

\begin{abstract}
Any two-qubit state can be represented geometrically by a steering ellipsoid inside the Bloch sphere. We extend this approach to represent any block positive two-qubit operator $B$. We derive a classification scheme based on the positivity of $\det B$ and $\det B^\mathrm{T_B}$; this shows that any ellipsoid inside the Bloch sphere must represent either a two-qubit state or a two-qubit entanglement witness. We focus on such witnesses and their corresponding ellipsoids, finding that properties such as witness optimality are naturally manifest in this geometric representation.\end{abstract}

\maketitle

\section{Introduction}

The characterisation, classification and detection of entanglement in a mixed quantum state constitute a fundamental open problem in quantum information theory. Entanglement witnesses provide one important approach to this problem~\cite{Guhne2009}. An entanglement witness~\cite{TerhalEW} is an operator that detects the presence of entanglement through the expectation value of an observable; any entangled state can be detected using an appropriate witness. Experimentally, entanglement witnesses provide a method for characterising a quantum state without needing full tomographic knowledge of the system~\cite{Eisert2007}. Mathematically, the theory of entanglement witnesses gives a very nontrivial generalisation of positive semidefinite operators (for a recent review, see Ref.~\cite{Review}).

A system of two qubits is the most basic unit for quantum entanglement. For such a system the Peres-Horodecki criterion~\cite{PeresPH,HorodeckiPH} gives a simple necessary and sufficient condition for detecting entanglement. However, two-qubit entanglement witnesses are still of interest in a variety of scenarios such as secure quantum key distribution~\cite{EW4, EW4_2}, the investigation of Bell nonlocality~\cite{Hyllus2005}, and the experimental characterisation of polarisation-entangled photons~\cite{Park2010}.

The steering ellipsoid formalism~\cite{Shi2011,Verstraete2002,Jevtic2013,Altepeter09} gives a faithful representation of two-qubit states analogous to the Bloch vector picture for a single qubit. Any two-qubit state may be represented by an ellipsoid inside the Bloch sphere, but not all ellipsoids inside the Bloch sphere represent a two-qubit state. Ref.~\cite{Monogamy} gave necessary and sufficient conditions for an ellipsoid to represent a state. In this paper we extend the steering ellipsoid formalism to also represent two-qubit entanglement witnesses. We will see that an ellipsoid inside the Bloch sphere must represent either a state or an entanglement witness. This gives an elegant physical interpretation to \textit{all} ellipsoids inside the Bloch sphere. Refs.~\cite{Monogamy,Obesity} examined two-qubit states of particular significance in the steering ellipsoid picture; this paper examines two-qubit entanglement witnesses from a similar geometric perspective.

Section \ref{preliminaries} introduces the basic theory to show how two-qubit operators can be represented by ellipsoids. In Section \ref{characterising} we give determinant-based criteria for the characterisation of two-qubit operators. This leads to a classification scheme for all ellipsoids inside the Bloch sphere. Section \ref{sec_EW} then investigates two-qubit entanglement witnesses in more detail. We study how witness optimality is manifest in the ellipsoid representation and look at several important examples of two-qubit entanglement witnesses. Finally, we give a conjecture that relates the ellipsoid picture to the notion of optimality within a set of entanglement witnesses.

\section{Preliminaries}\label{preliminaries}

\subsection{States and entanglement witnesses}

We begin by reviewing some basic definitions and setting out the notation. Let $R$ be a Hermitian operator acting on the Hilbert space $\mathcal{H}$. $R$ is \textit{positive semidefinite} ($R\geq 0$) when $\bra{\psi}R\ket{\psi}\geq 0$ for all $\ket{\psi}\in\mathcal{H}$. To be a quantum state we also require that $R$ has unit trace. $R$ is \textit{block positive} when $\bra{\psi}R\ket{\psi}\geq 0$ for all product $\ket{\psi}=\ket{\phi}\otimes\ket{\nu}\in\mathcal{H}$. An entanglement witness is block positive but not positive semidefinite~\cite{Terhal2000} and can without loss of generality be taken to have unit trace~\cite{OptimizationEW}. We will denote a Hermitian operator $R$, a block positive operator $B$, a state $\rho$, and an entanglement witness $W$. All of these will be unit trace operators acting on $\mathcal{H}=\mathbb{C}^2\otimes\mathbb{C}^2$. 

\subsection{Canonical two-qubit operators}

Consider a Hermitian operator $R$; in the Pauli basis $\{\I, \vec \sigma\}^{\otimes 2}$ we have\begin{equation}\label{eq:general_R}
R=\frac{1}{4}(\I\otimes\I+\vec a\cdot\vec \sigma\otimes\I+\I\otimes\vec b \cdot\vec \sigma+\sum_{i,j=1}^3T_{ij}\,\sigma_i\otimes\sigma_j),
\end{equation}
where $\vec a=\tr(R\,\vec \sigma \otimes \I)$, $\vec b=\tr(R\,\I\otimes \vec \sigma)$ and $T_{ij}=\tr(R \,\sigma_i \otimes \sigma_j)$. Note that $R$ is unit trace by construction but may or may not be positive semidefinite or block positive.

When $R$ is a state shared between Alice and Bob, we can represent it using Alice's steering ellipsoid $\E$. This gives the set of Bloch vectors to which Alice's qubit can be collapsed given all possible local measurements by Bob~\cite{Verstraete2002,Jevtic2013}. $\E$ necessarily lies inside the Bloch sphere~\footnote{Slightly abusing notation, we use the term `Bloch sphere' to mean the unit ball with centre $\vec 0$. Thus by `inside' we include the possibility that $\E$ is the whole Bloch sphere.}.  $\E$ is defined by its centre vector $\vec c$ and a real $3\times 3$ ellipsoid matrix $Q$~\cite{Jevtic2013}:
\begin{align}
\vec c &=\gamma_b^{2}(\vec a - T \vec b),\label{eq:centre}\\
Q &=\gamma_b^{2}\left(T-\vec a \vec b^\mathrm{T} \right)\left(\I + \gamma_b^2 \vec b \vec b^\mathrm{T}\right)\left(T^\mathrm{T}-\vec b \vec a^\mathrm{T} \right),\label{eq:Qmatrix}
\end{align}
where $\gamma_b=1/\sqrt{1-b^2}$ and $b=|\vec b|$. The eigenvalues of $Q$ are the squares of the ellipsoid semiaxes and the eigenvectors give the orientation of these axes. Note that $\E$ could be a degenerate ellipsoid, i.e. an ellipse, line or point, corresponding to rank deficient $Q$.

Although $\E$ is a steering ellipsoid only for the case that $R$ is a state, we can  define an ellipsoid in this way for any two-qubit operator of the form \eqref{eq:general_R}. Thus $\E$ will always be defined by its centre $\vec c$ and ellipsoid matrix $Q$, as given by \eqref{eq:centre} and \eqref{eq:Qmatrix}. $\E$ for an arbitrary $R$ will not necessarily lie inside the Bloch sphere.

As in Refs.~\cite{Shi2011,Jevtic2013,Monogamy,Obesity} we perform a reversible, trace-preserving local filtering operation that transforms $R$ to a canonical operator $\wt{R}$: 
\begin{equation}\label{eq:filtering}
\wt R=\left(\I\otimes \frac{1}{\sqrt{ 2R_B}}\right)\,R\,\left(\I\otimes\frac{1}{\sqrt{ 2R_B}}\right),
\end{equation}
where $R_B=\tr_A R$~\footnote{When $R_B$ is singular ($b=1$) the canonical transformation cannot be performed; \eqref{eq:centre} and \eqref{eq:Qmatrix} do not then apply, and $\E$ is simply a point at $\vec a$. This corresponds to $R$ being a product state (given that $\E$ lies within the Bloch sphere).}. In this canonical frame $\vec{ \wt b}=\vec 0$ and $\vec{ \wt a} = \vec c$ so
\begin{equation}\label{eq:canonical}
\wt R=\frac{1}{4}(\I\otimes\I+\vec c \cdot\vec \sigma\otimes\I+\sum_{i,j=1}^3 \wt{T}_{ij}\,\sigma_i\otimes\sigma_j).
\end{equation}
The ellipsoid matrix is defined in terms of this canonical operator as $Q=\wt{T}\wt{T}^\mathrm{T}$. The canonical operator is also used for defining the chirality of $\E$ as $\chi=\mathrm{sign}(\det \wt T)$~\cite{Monogamy}. We refer to an ellipsoid with $\chi=+1$ as \textit{right-handed} and an ellipsoid with $\chi=-1$ as \textit{left-handed}. A degenerate ellipsoid corresponds to $\chi=0$. The chirality of a steering ellipsoid has implications for the entanglement of a two-qubit state~\cite{Monogamy}, and we will see that it is also an important notion when characterising a general $\E$.

Crucially $\E$ is invariant under the canonical transformation \eqref{eq:filtering}. Local filtering operations also maintain positivity and block positivity~\cite{Avron2007}. Consequently, $R$ is a state if and only if $\wt R$ is a state, and $R$ is an entanglement witness if and only if $\wt R$ is an entanglement witness. This means that to characterise any ellipsoid describing a general two-qubit operator $R$ we need consider only the canonical operator $\wt R$.

\section{Characterising two-qubit operators using ellipsoids}\label{characterising}

\subsection{Block positivity}\label{sec_block_pos}

Although the ellipsoid $\E$ is defined for any Hermitian, unit trace two-qubit operator $R$, block positive operators have a particular significance.

\begin{theorem}\label{block_positive}Let $R$ be a two-qubit operator represented by ellipsoid $\E$. $R$ is block positive if and only if $\E$ lies inside the Bloch sphere.
\begin{proof}Since $\E$ and block positivity of $R$ are both invariant under the transformation \eqref{eq:filtering}, it suffices to consider a canonical operator $\wt R$ of the form \eqref{eq:canonical}. $\wt R$ is block positive when $\bra{\psi}\wt R\ket{\psi}\geq 0$ for all product $\ket{\psi}=\ket{\phi}\otimes\ket{\nu}$. Let $\vec \phi=\bra{\phi}\vec \sigma\ket{\phi}$ and $\vec \nu=\bra{\nu}\vec \sigma\ket{\nu}$ be the Bloch vectors, where we must have $|\vec \phi|=|\vec \nu|=1$. Then
$\bra{\psi}\wt R\ket{\psi}=\frac{1}{4}(1+\vec\phi\cdot\vec r^{(\vec \nu)})$, where $\vec r^{(\vec \nu)}$ has components $r^{(\vec \nu)}_i=c_i+\sum_{j=1}^3 \wt{T}_{ij} \nu_j$. Since $|\vec \nu|=1$, this describes the linear transformation of a unit sphere and in fact gives the ellipsoid $\E$ as defined in \eqref{eq:centre} and \eqref{eq:Qmatrix}~\cite{Jevtic2013}. $\vec r^{(\vec \nu)}$ is therefore a point on the surface of $\E$, parametrised by $\vec \nu$.

So $\bra{\psi}\wt R\ket{\psi}\geq 0$ for all $\ket{\psi}=\ket{\phi}\otimes\ket{\nu}$ if and only if $\vec\phi\cdot\vec r^{(\vec \nu)}\geq -1$ for all $|\vec \phi|=|\vec \nu|=1$.  This inequality is satisfied if and only if $|\vec r^{(\vec \nu)}|\leq 1$ for all $\vec \nu$, i.e. if and only if every point on $\E$ lies within the unit sphere.
\end{proof}
\end{theorem}

It should be noted that determining whether a general $\E$ lies inside the Bloch sphere is a difficult problem in Euclidean geometry~\cite{MathOverflow}. In fact, given Theorem \ref{block_positive}, the problem is clearly equivalent in difficulty to determining whether $R$ is block positive. This is known to be a hard problem, and there is no straightforward test that gives necessary and sufficient conditions for block positivity even in this simplest case of a $4\times 4$ matrix~\cite{BlockP}. In the Choi-isomorphic setting, the question is equivalent to determining whether a single-qubit map is positive, which is again known to be a hard problem (see, for example, Refs.~\cite{Pmap1, Pmap2}). However, often it will be plainly apparent whether $\E$ lies inside the Bloch sphere from a visualisation, and hence it will be immediately possible to determine block positivity of $R$ from the ellipsoid representation.

\subsection{Determinant criteria for states and entanglement witnesses}\label{det_criteria}

We now present a novel way of characterising two-qubit block positive operators $B$. This allows states and entanglement witnesses to be distinguished based on the positivity of the determinant alone.

\begin{theorem}\label{state_witness}Let $B$ be a two-qubit block positive operator. 
Then $B$ is a state if and only if $\det B\geq 0$; otherwise $B$ is an entanglement witness.
\begin{proof}
By definition $B$ is a state when $B\geq 0$. Since $B$ is block positive, by definition $B$ is an entanglement witness when $B\not\geq 0$. Clearly an operator $B\geq 0$ achieves $\det B\geq 0$. A two-qubit entanglement witness $B$ must have exactly one negative and three positive eigenvalues~\cite{Sarbicki2008} and hence $\det B<0$. The condition $\det B\geq 0$ is therefore necessary and sufficient for $B\geq 0$.\end{proof}
\end{theorem}

Note that the partially transposed operator $\BTB$ is block positive if and only if $B$ is block positive. It is already known that a two-qubit state $B$ is entangled if and only if $\det \BTB<0$~\cite{Jevtic2013,detrhoref}. Using Theorem \ref{state_witness}, the positivity of $\det B$ and $\det \BTB$ can then be used to classify all block positive two-qubit operators. For convenience we label these Classes A, B, C and D. Note that an operator $B$ belonging to Class B is equivalent to the operator $\BTB$ belonging to Class C:

\begin{widetext}\begin{align*}B \text{ and } \BTB \text{ are separable states}
&\Longleftrightarrow\det B\geq 0\text{ and } \det B^\mathrm{T_B}\geq 0\hspace{10mm}&\text{(Class A)}\\
\begin{rcases}
B \text{ is an entangled state } \\
B^\mathrm{T_B} \text{ is an entanglement witness } \end{rcases}
&\Longleftrightarrow\det B\geq 0\text{ and } \det B^\mathrm{T_B}< 0\hspace{10mm}&\text{(Class B)}\\
\begin{rcases}
B \text{ is an entanglement witness }   \\
B^\mathrm{T_B} \text{ is an entangled state }  \end{rcases}
&\Longleftrightarrow\det B< 0\text{ and } \det B^\mathrm{T_B}\geq0\hspace{10mm}&\text{(Class C)}\\
B \text{ and } \BTB \text{ are entanglement witnesses}
&\Longleftrightarrow\det B< 0\text{ and } \det B^\mathrm{T_B}<0\hspace{10mm}&\text{(Class D)}
\end{align*}
\end{widetext}

\subsection{Classifying block positive operator ellipsoids}\label{subsec_classifyingB}

\begin{figure*}
\subfigure[\, Class A]{\includegraphics[width=0.25\linewidth]{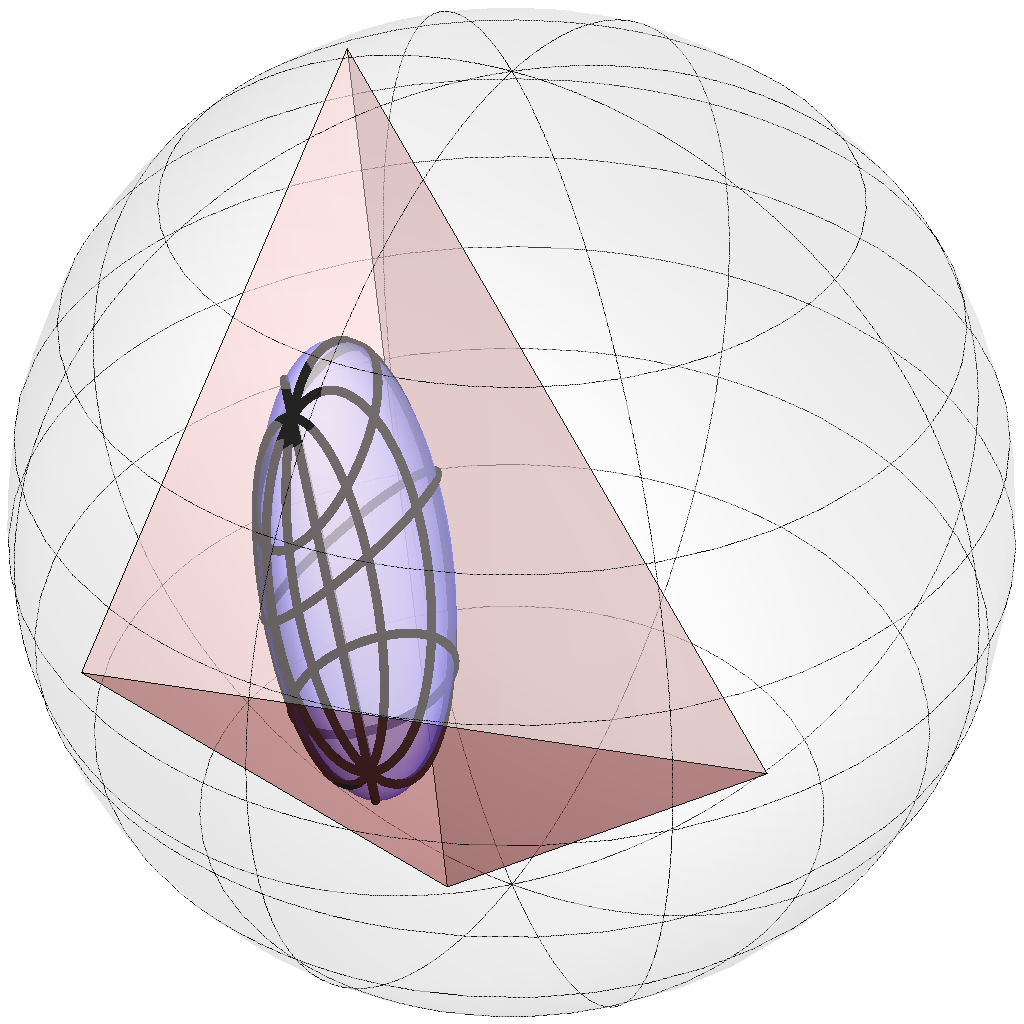}}\hspace{8mm}
\subfigure[\, Class B and C]{\includegraphics[width=0.25\linewidth]{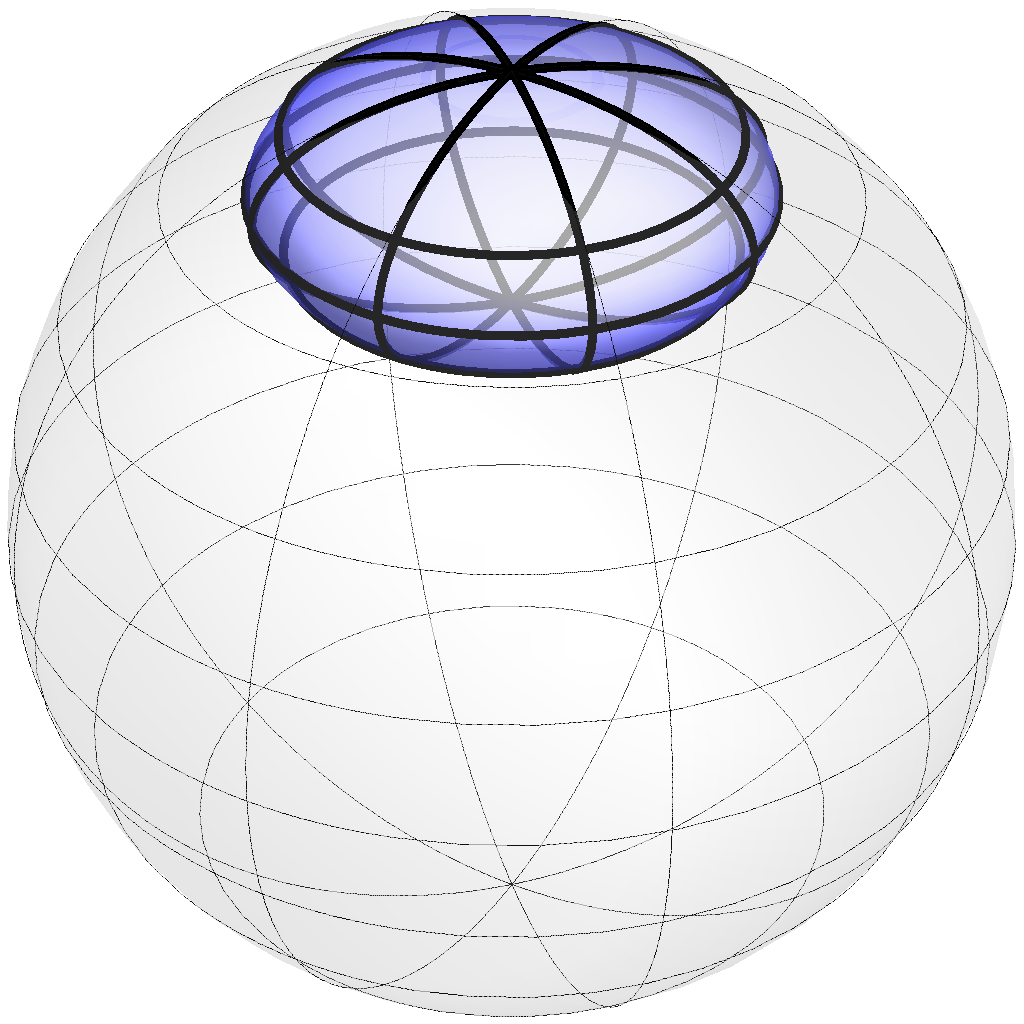}}\hspace{8mm}
\subfigure[\, Class D]{\includegraphics[width=0.25\linewidth]{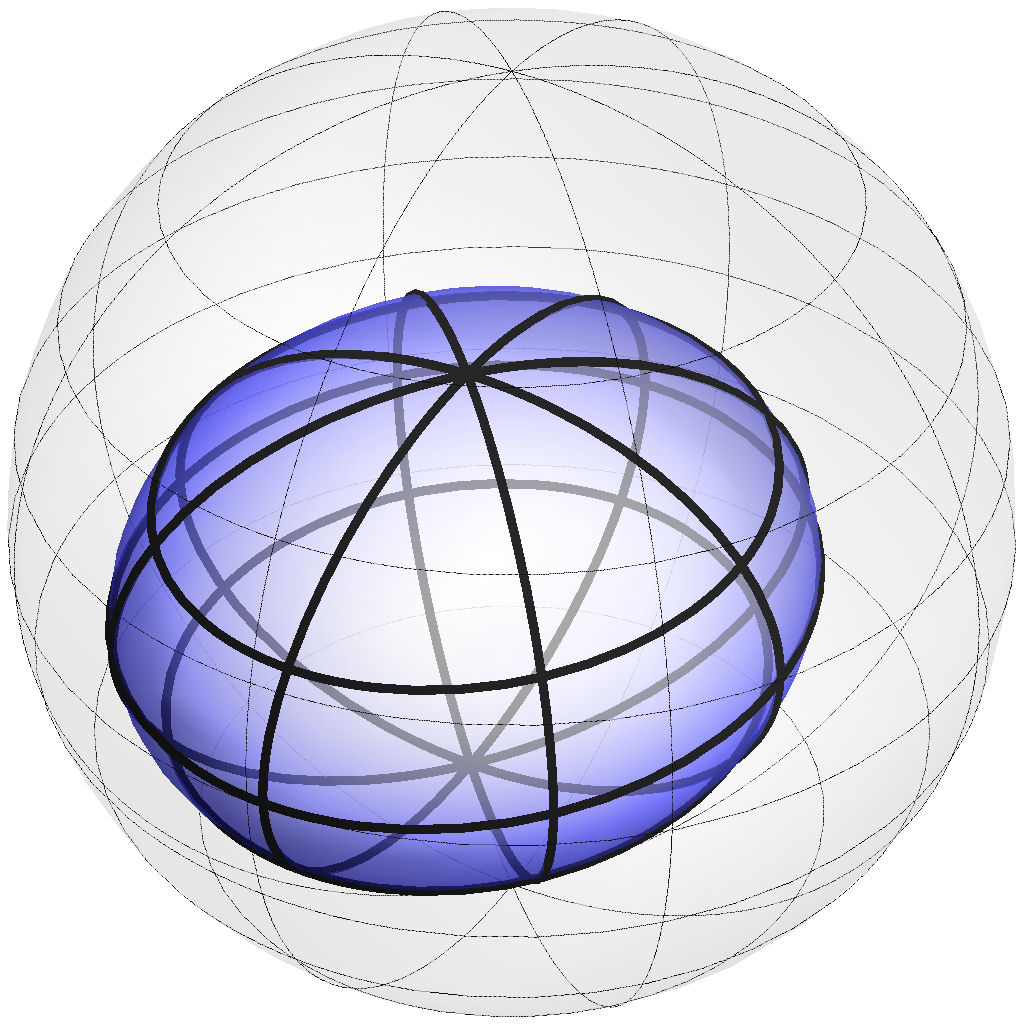}}

\caption{Visualisation of example $\E$ belonging to the different Classes given in Theorem \ref{classify_E}. (i) $\E$ belongs to Class A if and only if it fits inside a tetrahedron inside the Bloch sphere. Both the left- and right-handed $\E$ are separable states. (Image source: Ref.~\cite{Jevtic2013}.) (ii) The same surface describes $\E$ belonging to Class B and $\E$ belonging to Class C. The left-handed $\E$ represents an entangled state; the right-handed $\E$ represents an entanglement witness. Here we show the largest volume $\E$ that fits inside the Bloch sphere for fixed ellipsoid centre $\vec c=(0,0,\frac{1}{2})$~\cite{Obesity}. (iii) $\E$ belonging to Class D represents an entanglement witness in both its left- and right-handed forms.}
\label{fig:classes}
\end{figure*}

Due to Theorem \ref{block_positive}, any $\E$ inside the Bloch sphere describes a block positive two-qubit operator $B$ and therefore can be classified using the scheme presented above. Recall that the canonical transformation \eqref{eq:filtering} maintains positivity and block positivity. This means that for block positive $B$ we have $\det B\geq 0\Leftrightarrow\det \wt B\geq 0$ and $\det \BTB\geq 0\Leftrightarrow\det \wtBTB\geq 0$. Since expressions involving $\wt B$ can be written in terms of the ellipsoid centre $\vec c$, matrix $Q$ and chirality $\chi$, this allows us to characterise \textit{any} block positive two-qubit operator using geometric features of $\E$. 

Expressions for $\det \wt B$ and $\det \wtBTB$ were found in Ref.~\cite{Monogamy}. Since partial transposition is equivalent to flipping the ellipsoid chirality ($\chi \rightarrow -\chi$), these are identical apart from the sign of one term:
\begin{align}
\det \wt B \geq 0  & \Leftrightarrow c^4-2 u c^2+q-\chi r\geq 0,\label{detR}\\
\det \wtBTB \geq 0 & \Leftrightarrow c^4-2 u c^2+q+\chi r\geq 0,\label{detRTB}
\end{align}
where
\begin{align}u&=1-\tr Q+2\vec{\hat c}^\mathrm{T}Q\vec{\hat c},\label{eq:u} \\
q&=1+2\tr(Q^2)-2\tr Q - (\tr Q)^2, \label{eq:q}\\
r&=8 \sqrt{\det Q},\label{eq:r}\end{align}
with the unit vector $\vec {\hat c}=\vec c/c$.

\begin{theorem}\label{classify_E}Let $\E$ be an ellipsoid lying inside the Bloch sphere, with centre $\vec c$, matrix $Q$ and chirality $\chi$. The block positive two-qubit operator $B$ that is represented by $\E$ can be classified according to the ellipsoid parameters:
\begin{widetext}\begin{align*}B \text{ and } \BTB \text{ are separable states}
&\Longleftrightarrow c^4-2 u c^2+q- r\geq 0\hspace{10mm}&\text{(Class A)}\\
\begin{rcases}
B \text{ is an entangled state } \\
B^\mathrm{T_B} \text{ is an entanglement witness } \end{rcases}
&\Longleftrightarrow
\begin{cases}c^4-2 u c^2+q-\chi r\geq 0 \\
c^4-2 u c^2+q+\chi r< 0
\end{cases}
\hspace{10mm}&\text{(Class B)}\\
\begin{rcases}
B \text{ is an entanglement witness }   \\
B^\mathrm{T_B} \text{ is an entangled state }  \end{rcases}
&\Longleftrightarrow
\begin{cases}c^4-2 u c^2+q-\chi r< 0 \\
c^4-2 u c^2+q+\chi r\geq 0
\end{cases}
\hspace{10mm}&\text{(Class C)}\\
B \text{ and } \BTB \text{ are entanglement witnesses}
&\Longleftrightarrow c^4-2 u c^2+q+ r<0&\text{(Class D)}
\end{align*}\end{widetext}
where $u$, $q$ and $r$ are given by \eqref{eq:u}, \eqref{eq:q} and \eqref{eq:r}. 

\hspace{100mm}
\vspace{100mm}

\begin{proof}Since $\det B\geq 0\Leftrightarrow\det \wt B\geq 0$ and $\det \BTB\geq 0\Leftrightarrow\det \wtBTB\geq 0$, we can directly convert the classification scheme given in Section \ref{det_criteria} and use \eqref{detR} and \eqref{detRTB}. The necessary and sufficient conditions for $\E$ to belong to Class A are therefore $c^4-2 u c^2+q-\chi r\geq 0$ and $c^4-2 u c^2+q+\chi r\geq 0$. These two inequalities are equivalent to $c^4-2 u c^2+q-|\chi r|\geq 0$. However, $r\geq 0$ and $\chi=\pm 1,0$ and so $|\chi r|=r$, where the case of a degenerate $\E$ also holds since $r = 0$ if and only if $\chi=0$. Hence the single inequality $c^4-2 u c^2+q- r\geq 0$ is necessary and sufficient for Class A. The two inequalities for Class D simplify similarly.\end{proof}
\end{theorem}

Any $\E$ inside the Bloch sphere can thus be straightforwardly classified according to its geometric features. As in Refs.~\cite{Jevtic2013} and \cite{Monogamy} we can identify three geometric contributions: the distance of the centre of $\E$ from the origin, the size of $\E$ (through terms such as $\tr Q$ and $\det Q$) and the skew $\vec{\hat c}^\mathrm{T}Q\vec{\hat c}$ (which gives a measure of the orientation of $\E$ relative to $\vec c$).

Fig. \ref{fig:classes} shows example ellipsoids for each Class. We now make a few remarks to highlight how Theorem \ref{classify_E} and the notion of ellipsoid chirality can be used together with geometric properties to classify an ellipsoid.

\begin{itemize}
\item As discussed in Ref.~\cite{Monogamy}, $\E$ for an entangled state (Class B) must be left-handed, as it obeys $\chi r < -\chi r$. We see similarly that $\E$ belonging to Class C must obey $\chi r > -\chi r$ and therefore be right-handed.
\item Any degenerate $\E$ inside the Bloch sphere must belong to Class A or Class D. The \textit{nested tetrahedron condition}~\cite{Jevtic2013} states that $\E$ fits inside a tetrahedron inside the Bloch sphere if and only if it corresponds to a separable state (Class A). For the case of degenerate $\E$, the nested tetrahedron may be taken to be a triangle; degenerate $\E$ belonging to Class D are therefore those which do not fit inside a triangle inside the Bloch sphere.
\item Non-degenerate $\E$ belonging to Class A are those for which both the left- and right-handed ellipsoids represent separable states. Non-degenerate $\E$ belonging to Class D are those for which both the left- and right-handed ellipsoids represent entanglement witnesses.
\item Any $\E$ that meets the surface of the Bloch sphere at a circle cannot represent a state regardless of its chirality~\cite{Monogamy,Braun2013}; such ellipsoids must therefore belong to Class D.
\end{itemize}

Ref.~\cite{Monogamy} gave necessary and sufficient conditions for a two-qubit operator to represent a state (separable or entangled). Theorem \ref{classify_E} gives an alternative formulation of this: given that $\E$ lies inside the Bloch sphere, it represents a state if and only if $\E$ belongs to Class A or Class B. Any $\E$ inside the Bloch sphere that does not represent a state must instead represent an entanglement witness (Class C or Class D). This gives a new physical interpretation to ellipsoids that were previously considered unphysical. In the remainder of this paper we will investigate these ellipsoids and the corresponding entanglement witnesses in more detail.

\section{Ellipsoids for two-qubit entanglement witnesses}\label{sec_EW}

\subsection{Definitions}

$W$ will denote a unit trace two-qubit entanglement witness, which could belong to either Class C or Class D. A state $\rho$ is detected by $W$ when $\tr (\rho W)<0$, and a witness $W_1$ is said to be \textit{finer} than another witness $W_2$ if all the states detected by $W_2$ are also detected by $W_1$. $W$ is called \textit{optimal} when there does not exist a finer witness~\cite{OptimizationEW}. This notion can be extended to optimality within a set~\cite{EW4}: let $S$ be a set of entanglement witnesses; $W \in S$ is optimal in $S$ if there does not exist a finer entanglement witness in $S$. Finally, $W$ is \textit{weakly optimal} when there exists a product state $\ket{\psi}=\ket{\phi}\otimes\ket{\nu}\in\mathcal{H}$ such that $\bra{\psi}W\ket{\psi}=0$~\cite{WeaklyOptimal}.

\subsection{Optimality and weak optimality}

The properties of optimality and weak optimality can be immediately visualised using the ellipsoid representation.

\begin{theorem}\label{optimality}Let $W$ be a two-qubit entanglement witness represented by ellipsoid $\E$. Then

\begin{enumerate}[label=(\roman*),noitemsep,nolistsep]
\item $W$ is optimal if and only if $\E$ is the whole Bloch sphere and right-handed;
\item $W$ is weakly optimal if and only if $\E$ touches the surface of the Bloch sphere.
\end{enumerate}
\begin{proof}$\,$
\begin{enumerate}[label=(\roman*),noitemsep,nolistsep]
\item An optimal two-qubit entanglement witness is of the form $W=\ket{\psi_e}\bra{\psi_e}^\mathrm{T_B}$, with $\ket{\psi_e}$ an entangled state. The steering ellipsoid for a state $\rho$ is the whole Bloch sphere if and only if $\rho$ is a pure entangled state~\cite{Jevtic2013}, and such a steering ellipsoid must be left-handed~\cite{Monogamy}. An optimal entanglement witness is the partial transposition of such a state ($\rho=\ket{\psi_e}\bra{\psi_e}$). Since partial transposition leaves the ellipsoid surface invariant but flips the chirality, this corresponds to the case that $\E$ is the whole Bloch sphere and right-handed.
\item That the property of weak optimality is preserved under the canonical transformation is clear from \eqref{eq:filtering}: there exists $\ket{\psi}=\ket{\phi}\otimes\ket{\nu}$ such that $\bra{\psi}\wt W\ket{\psi}=0$ if and only if there exists  $\ket{\psi'}=\ket{\phi}\otimes\ket{\nu'}$  such that $\bra{\psi'}W\ket{\psi'}=0$. Since $\E$ is invariant under the canonical transformation, it therefore suffices to consider a canonical entanglement witness $\wt W$. The proof then proceeds similarly to Theorem \ref{block_positive}. There exists $\ket{\psi}=\ket{\phi}\otimes\ket{\nu}$ such that $\bra{\psi}\wt W\ket{\psi}=0$ if and only if there exists $\vec \phi$ with $|\vec \phi|=1$ such that $\vec\phi\cdot\vec r^{(\vec \nu)}= -1$ for some point $\vec r^{(\vec \nu)}$ on the surface of $\E$ parametrised by $\vec \nu$. Clearly $\vec\phi\cdot\vec r^{(\vec \nu)}= -1$ implies that there exists some $\vec \nu$ for which $|\vec r^{(\vec \nu)}|=1$. Conversely, since the only constraint on $\vec \phi$ is $|\vec \phi|=1$, if $|\vec r^{(\vec \nu)}|=1$ then the direction of $\vec \phi$ can always be chosen so $\vec\phi\cdot\vec r^{(\vec \nu)}= -1$. So $\wt W$ is weakly optimal if and only if $|\vec r^{(\vec \nu)}|=1$, i.e. a point on $\E$ touches the surface of the Bloch sphere.\end{enumerate}
\end{proof}
\end{theorem}

\begin{figure}
\includegraphics[width=\columnwidth]{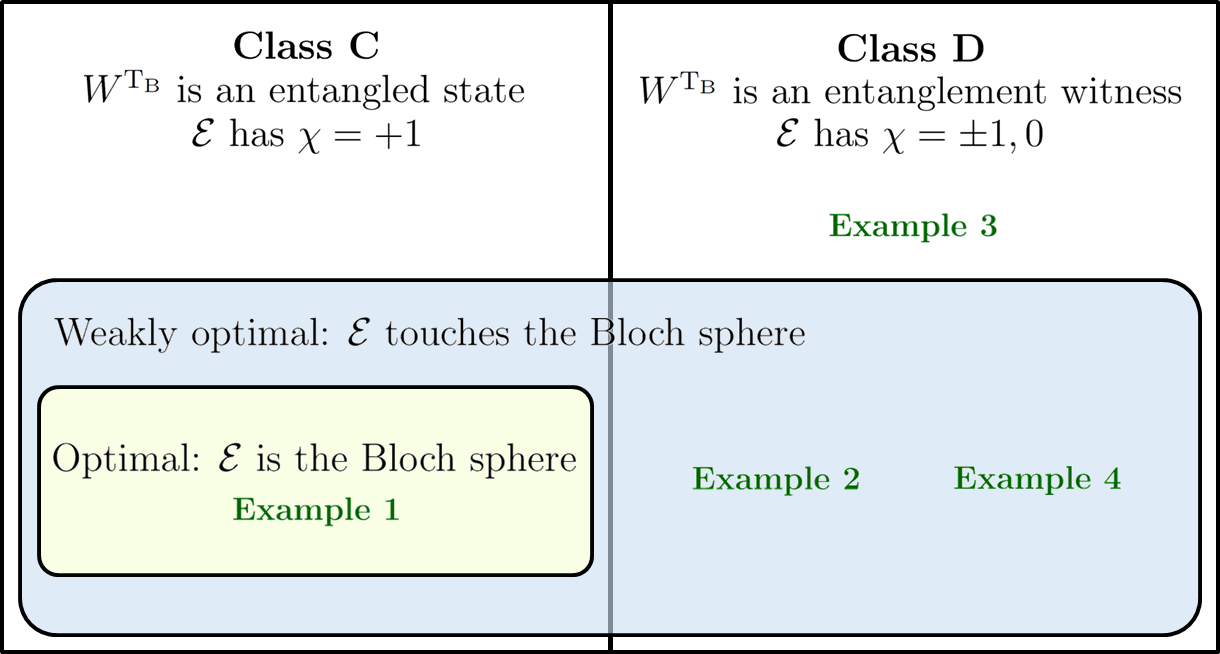}

\caption{Any two-qubit entanglement witness $W$ belongs to either Class C or Class D; these are distinguished by whether $W^\mathrm{T_B}$ is an entangled state or an entanglement witness. Witnesses represented by degenerate $\E$ must belong to Class D, while all witnesses in Class C must be right-handed. There are weakly optimal witnesses in Class C and Class D, but an optimal witness must belong to Class C. The optimality or weak optimality of a witness is immediately evident from a visualisation of $\E$ inside the Bloch sphere. Example witnesses discussed in the main text are shown.}
\label{fig:class_C_vs_D}
\end{figure}

We thus see from a geometric perspective that an optimal witness is a special case of a weakly optimal witness. In terms of the classification scheme given in Theorem \ref{classify_E}, optimal $W$ must belong to Class C, since $W^\mathrm{T_B}=\ket{\psi_e}\bra{\psi_e}$ is an entangled state. A weakly optimal $W$ can belong to Class C or Class D (Fig. \ref{fig:class_C_vs_D}). 

\subsection{Examples of entanglement witnesses}

\begin{figure*}
\subfigure[\, Example 1]{\includegraphics[width=0.21\linewidth]{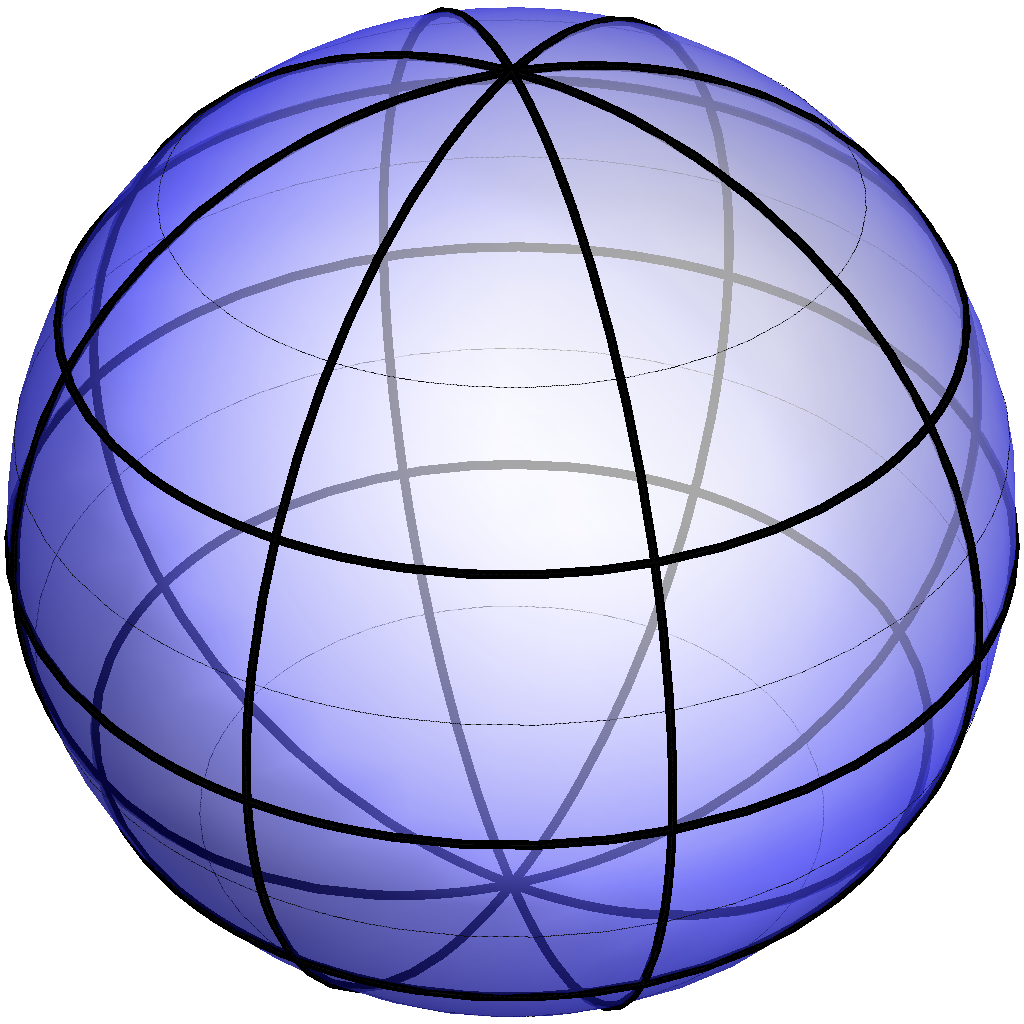}}\hspace{8mm}
\subfigure[\, Example 2]{\includegraphics[width=0.21\linewidth]{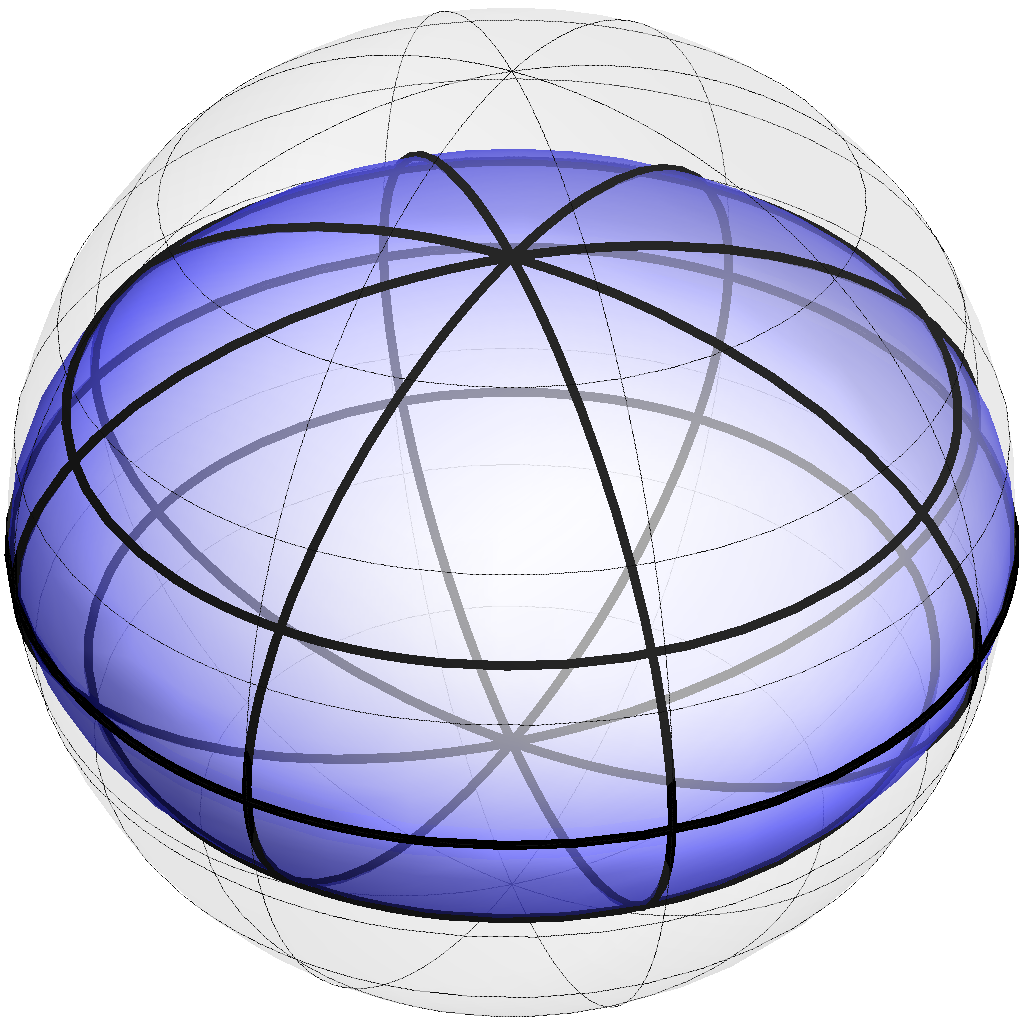}}\hspace{8mm}
\subfigure[\, Example 3]{\includegraphics[width=0.21\linewidth]{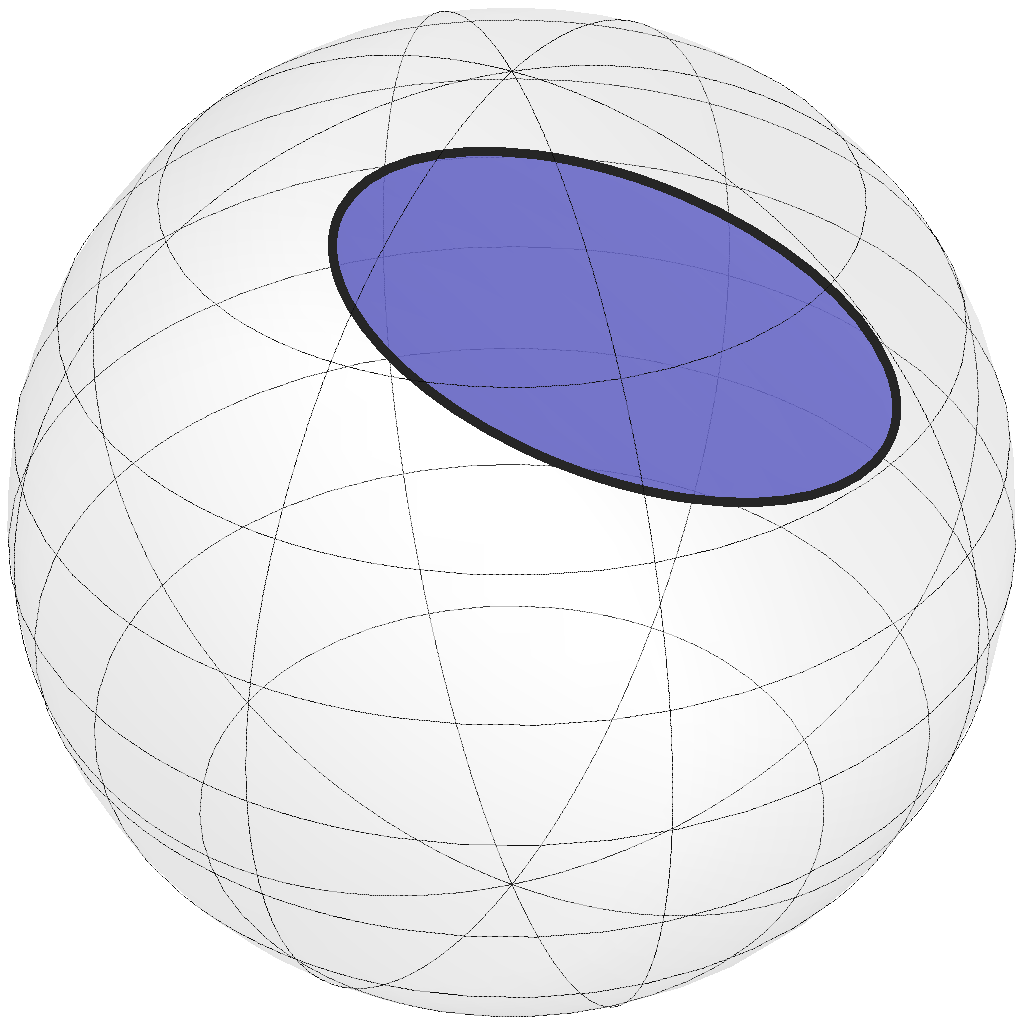}}\hspace{8mm}
\subfigure[\, Example 4]{\includegraphics[width=0.21\linewidth]{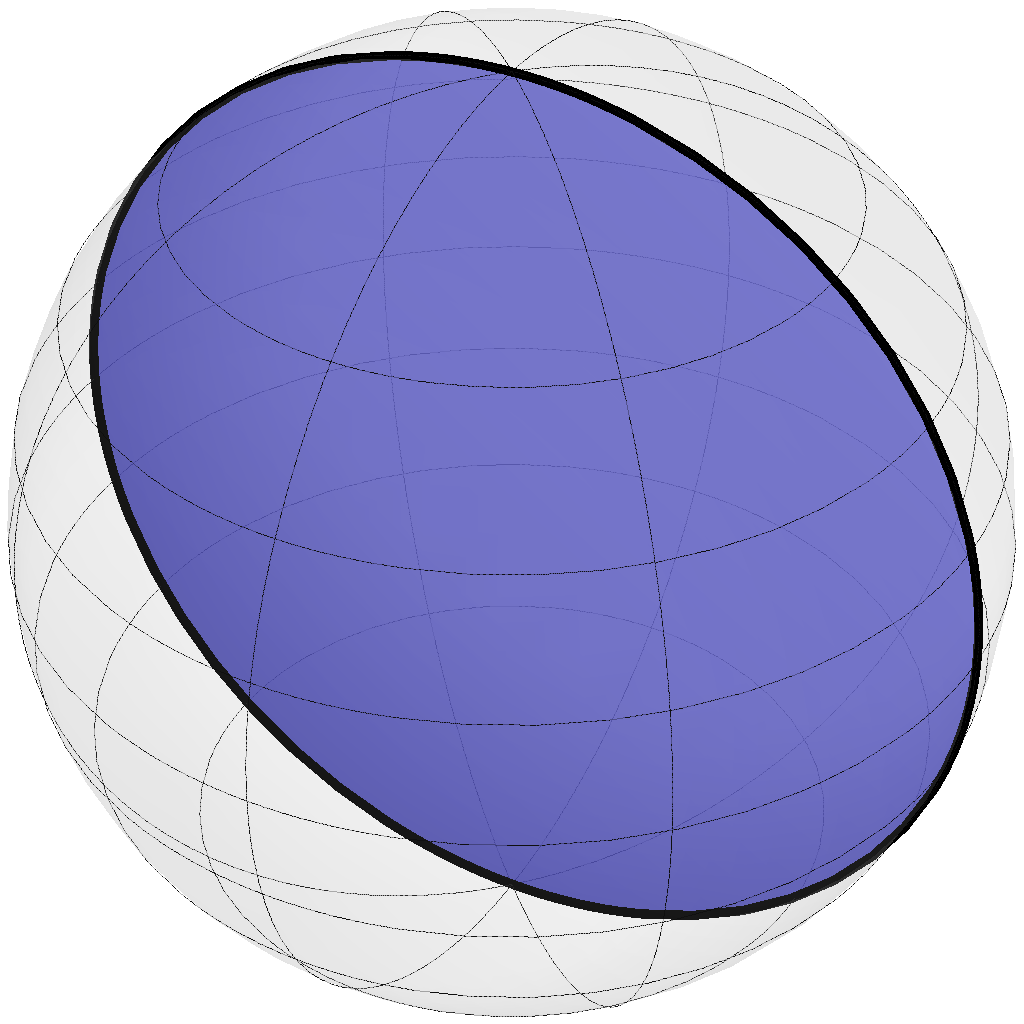}}

\caption{Visualisation of the example $\E$. (i) $\E$ representing the optimal witness $W=\ket{\phi^+}\bra{\phi^+}^\mathrm{T_B}$ is the whole Bloch sphere with $\chi=+1$. (ii) $\E$ representing a weakly optimal witness touches the surface of Bloch sphere. Here we show $\E$ for $W_p$ with $p=\frac{1}{5}$; this meets the surface of the Bloch sphere at the circle on the equatorial plane. (iii) $\E$ representing $W\in\EW4$ is an ellipse in the $xz$ plane. Due to the nested tetrahedron condition, there is no triangle inside the Bloch sphere that circumscribes $\E$. (iv) $\E$ representing optimal $W\in\EW4$ is the $xz$ unit disc.}
\label{fig:examples}
\end{figure*}

We now look at some examples of two-qubit entanglement witnesses to see how the geometric features of $\E$ relate to witness properties. This will also serve to illustrate the distinction between Class C and Class D (recall that $W$ belongs to Class C when $W^\mathrm{T_B}$ is an entangled state; $W$ belongs to Class D when $W^\mathrm{T_B}$ is an entanglement witness). The examples are shown on Fig. \ref{fig:class_C_vs_D}, with the corresponding ellipsoids visualised in Fig. \ref{fig:examples}.

\begin{example}\label{flip}The flip operator $\mathbb{F}$ is defined as $\mathbb{F}\ket{\phi}\otimes\ket{\nu}=\ket{\nu}\otimes\ket{\phi}$~\cite{Review}. After normalisation to unit trace we have $W=\frac{1}{2}\mathbb{F}=\ket{\phi^+}\bra{\phi^+}^\mathrm{T_B}$, where $\ket{\phi^+}=\frac{1}{\sqrt 2}(\ket{00}+\ket{11})$. In terms of the Pauli basis \eqref{eq:general_R}, we have $\vec a = \vec b =\vec 0$ and $T=\mathrm{diag}(1, 1, 1)$. From \eqref{eq:centre} and \eqref{eq:Qmatrix} we see that $\E$ representing $W$ is the whole Bloch sphere with $\chi=+1$, as it must be for an optimal entanglement witness (Theorem \ref{optimality}). As with any optimal entanglement witness, $W$ belongs to Class C.\end{example}

\begin{example}We consider a family of weakly optimal witnesses originally presented in Ref.~\cite{Korbicz2008} and studied further in Refs.~\cite{Wang2014,Wang2013}. After normalisation, the family may be parametrised by $p$ as
\begin{equation}\nonumber W_p=\frac{1}{2}\begin{pmatrix}
p & 0 & 0 & 0  \\
0 & 1-p & 1 & 0 \\
0 & 1 & 1-p & 0 \\
0 & 0 & 0 & p
\end{pmatrix}.
\end{equation}

In terms of the Pauli basis \eqref{eq:general_R}, we have $\vec a = \vec b =\vec 0$ and $T=\mathrm{diag}(1, 1, 2p-1)$. ${\E}_p$ representing $W_p$ therefore has centre $\vec c=\vec 0$ and chirality $\chi=\mathrm{sign}(2p-1)$. The semiaxes of ${\E}_p$ are length 1, 1 and $|2p-1|$ aligned with the $x$, $y$ and $z$ coordinate axes respectively.

${\E}_p$ lies inside the Bloch sphere if and only if $|2p-1|\leq 1$, so $W_p$ is block positive if and only if $0\leq p\leq 1$. $W_p$ is positive semidefinite for $p=0$, and so $W_p$ is an entanglement witness if and only if $0< p\leq 1$. Since all such ${\E}_p$ touch the surface of the Bloch sphere, $W_p$ forms a family of weakly optimal entanglement witnesses. (It is straightforward to verify that $\bra{\psi}W_p\ket{\psi}=0$ for $\ket{\psi}=\ket{+}\otimes\ket{-}$ with $\ket{\pm}=\frac{1}{\sqrt 2}(\ket{0}\pm\ket{1})$, fulfilling the defining criterion of a weakly optimal witness.) When $p=1$, $W_p$ reduces to the optimal witness presented in Example \ref{flip}. For all other $p$, the ellipsoid ${\E}_p$ meets the surface of the Bloch sphere at the circle on the equatorial plane. As discussed in Section \ref{subsec_classifyingB}, such ellipsoids belong to Class D.\end{example}

\begin{example}\label{eg_EW4}Refs.~\cite{EW4, EW4_2} introduce $\mathrm{EW}_4$, a set of two-qubit entanglement witnesses that is of interest in quantum key distribution. An entanglement witness $W\in \EW4$ if and only if $W=W^\mathrm{T}=W^\mathrm{T_B}$.

Using the Pauli basis expansion \eqref{eq:general_R}, for $W\in\EW4$ all terms involving $\sigma_2$ must vanish so
\begin{equation}\nonumber\vec a=\begin{pmatrix}a_1\\ 0\\ a_3\end{pmatrix}, \vec b=\begin{pmatrix}b_1\\0\\b_3\end{pmatrix} \text{ and }
 T=\begin{pmatrix}T_{11} & 0 & T_{13} \\ 0 & 0 & 0 \\ T_{31} & 0 & T_{33}\end{pmatrix}.\end{equation}
We then use \eqref{eq:centre} and \eqref{eq:Qmatrix} to find the corresponding ellipsoid $\E$. This has centre $\vec c$ lying in the $xz$ plane. The ellipsoid matrix $Q$ is rank deficient and so $\E$ is degenerate ($\chi=0$). The support of $Q$ spans the $xz$ plane and hence $\E$ itself must lie within the $xz$ plane. $\E$ cannot be a line or point, as these always describe a separable state (since they correspond to degenerate tetrahedra inside the Bloch sphere and hence satisfy the nested tetrahedron condition~\cite{Jevtic2013}). Therefore $\E$ for $W\in\EW4$ is an ellipse in the $xz$ plane. As a degenerate ellipsoid, all $\E$ for $W\in\EW4$ belong to Class D.\end{example}

\begin{example}\label{eg_optimal_EW4}
Entanglement witnesses that are optimal within the set $\EW4$ are given by $W=\frac{1}{2}(\rho+\rho^\mathrm{T_B})$, where $\rho=\ket{\psi_e}\bra{\psi_e}$ and $\ket{\psi_e}$ is a real entangled state~\cite{EW4}. Ref.~\cite{Ried2014} shows that $\E$ for such an operator is a circular disc with centre $\vec c=\vec 0$ and radius 1. Any $W\in\EW4$ must lie in the $xz$ plane, and so optimal witnesses within $\EW4$ are represented by the $xz$ unit disc itself. Note that these witnesses are also weakly optimal for two qubits in general, since $\E$ touches the surface of the Bloch sphere.\end{example}

\subsection{Optimality within a set: a conjecture}

The examples given above suggest an interesting new geometric perspective on optimality within a set of two-qubit entanglement witnesses. Consider the set $S$ of all two-qubit entanglement witnesses. The ellipsoids describing $W\in S$ always lie within the Bloch sphere, and the optimal $W \in S$ are simply the optimal two-qubit entanglement witnesses. According to Theorem \ref{optimality}, the ellipsoid representing these optimal $W$ is the whole Bloch sphere. This $\E$ is the largest possible one representing any $W\in S$.

Members of the set $\EW4$ are described by $\E$ that are ellipses within the $xz$ plane (Example \ref{eg_EW4}). The optimal $W\in\EW4$ are described by the whole $xz$ unit disc (Example \ref{eg_optimal_EW4}). Again, this $\E$ is the largest possible ellipsoid for any $W\in \EW4$. This leads us to conjecture that the optimal $W$ within a set will always be described by the largest possible ellipsoid.

\begin{conjecture*}Let $\E^\star$ be an ellipsoid lying inside the Bloch sphere and $S$ be a set of two-qubit entanglement witnesses defined as follows: $W\in S$ if and only if $\E$ representing $W$ lies inside $\E^\star$. The optimal $W\in S$ are then represented by $\E=\E^\star$.
\end{conjecture*}

In the case that $S$ is the set of all two-qubit entanglement witnesses, $\E^\star$ is the whole Bloch sphere; for our example $S=\EW4$, $\E^\star$ is the $xz$ plane. Note that this Conjecture applies to any $\E^\star$ belonging to Class C or Class D.

\begin{figure}
\includegraphics[width=0.6\columnwidth]{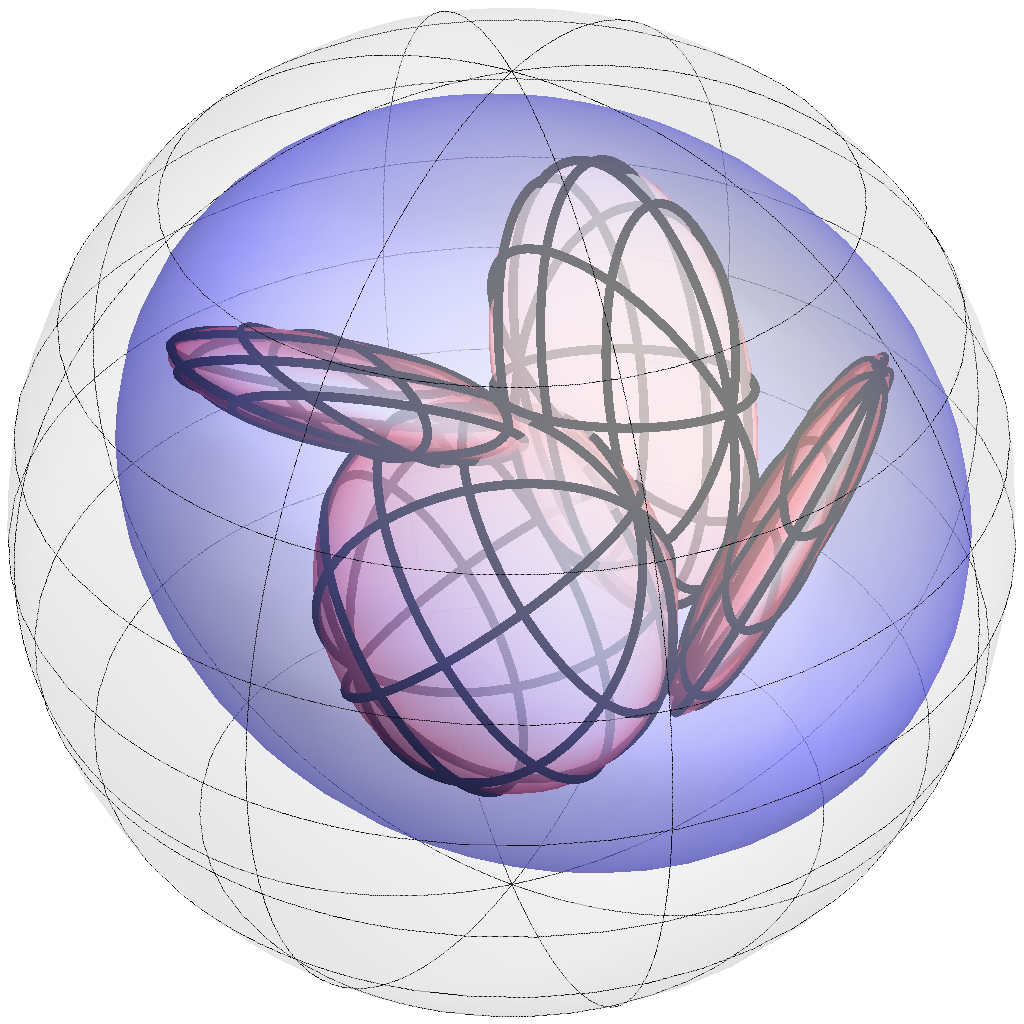}
\caption{Visualisation of the Conjecture. The blue ellipsoid is $\E^\star$; the red ellipsoids are example witnesses $W\in S$. We conjecture that the optimal $W\in S$ are described by $\E=\E^\star$.}
\label{fig:conjecture}
\end{figure}

Although this Conjecture is easy to visualise geometrically (Fig. \ref{fig:conjecture}), it is nontrivial to approach analytically. In addition to finding the optimal witnesses within a set, the Conjecture involves determining whether a given $W$ belongs to $S$. This means finding whether one ellipsoid $\E$ lies inside another $\E^\star$, which, as noted in Section \ref{sec_block_pos}, is a difficult problem.

\section{Conclusions}

The steering ellipsoid formalism for representing two-qubit states has been extended to represent any two-qubit block positive operator $B$. By classifying $B$ using the positivity of $\det B$ and $\det \BTB$, any ellipsoid inside the Bloch sphere can now be classified as a separable state, entangled state or entanglement witness. 

We have studied several examples of two-qubit entanglement witnesses and found that features such as optimality and weak optimality are clearly manifest in the ellipsoid representation. This promotes ellipsoids as a natural and intuitive scheme for representing two-qubit entanglement witnesses. The geometric view also leads to a new perspective on optimality within a set of two-qubit entanglement witnesses.

It is also worth noting some features of the formalism that we have considered without finding any significant results. Is there any relationship between the ellipsoid representing an entanglement witness $W$ and the set of ellipsoids describing two-qubit states that $W$ detects? Or, conversely, is there any relationship between the ellipsoid of an entangled two-qubit state and the set of ellipsoids describing entanglement witnesses which can detect that state? For a two-qubit state the steering ellipsoid represents the set of single-qubit states to which Alice can be steered given all possible local measurements by Bob; is there an analogous physical interpretation for what the entanglement witness ellipsoid itself represents? It also remains to be seen whether an analogous representation of entanglement witnesses is useful for studying higher-dimensional scenarios such as many-qubit systems.

Finally, we note that Wang et al.~\cite{Wang2014,Wang2013} have recently characterised weakly optimal entanglement witnesses and given a general procedure for their construction. The ellipsoid representation might give a novel geometric interpretation of this procedure for the case of two qubits.

\hspace{1mm}

\begin{acknowledgments}
We thank Dariusz Chru\'{s}ci\'{n}ski, Sania Jevtic and Matthew Pusey for useful discussions. This work was supported by EPSRC. DJ is funded by the Royal Society. TR is supported by the Leverhulme Trust.
\end{acknowledgments}

\bibliography{references}

\begin{thebibliography}{33}%
\makeatletter
\providecommand \@ifxundefined [1]{%
 \@ifx{#1\undefined}
}%
\providecommand \@ifnum [1]{%
 \ifnum #1\expandafter \@firstoftwo
 \else \expandafter \@secondoftwo
 \fi
}%
\providecommand \@ifx [1]{%
 \ifx #1\expandafter \@firstoftwo
 \else \expandafter \@secondoftwo
 \fi
}%
\providecommand \natexlab [1]{#1}%
\providecommand \enquote  [1]{``#1''}%
\providecommand \bibnamefont  [1]{#1}%
\providecommand \bibfnamefont [1]{#1}%
\providecommand \citenamefont [1]{#1}%
\providecommand \href@noop [0]{\@secondoftwo}%
\providecommand \href [0]{\begingroup \@sanitize@url \@href}%
\providecommand \@href[1]{\@@startlink{#1}\@@href}%
\providecommand \@@href[1]{\endgroup#1\@@endlink}%
\providecommand \@sanitize@url [0]{\catcode `\\12\catcode `\$12\catcode
  `\&12\catcode `\#12\catcode `\^12\catcode `\_12\catcode `\%12\relax}%
\providecommand \@@startlink[1]{}%
\providecommand \@@endlink[0]{}%
\providecommand \url  [0]{\begingroup\@sanitize@url \@url }%
\providecommand \@url [1]{\endgroup\@href {#1}{\urlprefix }}%
\providecommand \urlprefix  [0]{URL }%
\providecommand \Eprint [0]{\href }%
\providecommand \doibase [0]{http://dx.doi.org/}%
\providecommand \selectlanguage [0]{\@gobble}%
\providecommand \bibinfo  [0]{\@secondoftwo}%
\providecommand \bibfield  [0]{\@secondoftwo}%
\providecommand \translation [1]{[#1]}%
\providecommand \BibitemOpen [0]{}%
\providecommand \bibitemStop [0]{}%
\providecommand \bibitemNoStop [0]{.\EOS\space}%
\providecommand \EOS [0]{\spacefactor3000\relax}%
\providecommand \BibitemShut  [1]{\csname bibitem#1\endcsname}%
\let\auto@bib@innerbib\@empty
%</preamble>
\bibitem [{\citenamefont {G\"{u}hne}\ and\ \citenamefont
  {T\'oth}(2009)}]{Guhne2009}%
  \BibitemOpen
  \bibfield  {author} {\bibinfo {author} {\bibfnamefont {O.}~\bibnamefont
  {G\"{u}hne}}\ and\ \bibinfo {author} {\bibfnamefont {G.}~\bibnamefont
  {T\'oth}},\ }\href@noop {} {\bibfield  {journal} {\bibinfo  {journal} {Phys.
  Rep.}\ }\textbf {\bibinfo {volume} {474}},\ \bibinfo {pages} {1} (\bibinfo
  {year} {2009})}\BibitemShut {NoStop}%
\bibitem [{\citenamefont {Terhal}(2000)}]{TerhalEW}%
  \BibitemOpen
  \bibfield  {author} {\bibinfo {author} {\bibfnamefont {B.~M.}\ \bibnamefont
  {Terhal}},\ }\href@noop {} {\bibfield  {journal} {\bibinfo  {journal} {Phys.
  Lett. A}\ }\textbf {\bibinfo {volume} {271}},\ \bibinfo {pages} {319}
  (\bibinfo {year} {2000})}\BibitemShut {NoStop}%
\bibitem [{\citenamefont {Eisert}\ \emph {et~al.}(2007)\citenamefont {Eisert},
  \citenamefont {Brand\~{a}o},\ and\ \citenamefont {Audenaert}}]{Eisert2007}%
  \BibitemOpen
  \bibfield  {author} {\bibinfo {author} {\bibfnamefont {J.}~\bibnamefont
  {Eisert}}, \bibinfo {author} {\bibfnamefont {F.~G. S.~L.}\ \bibnamefont
  {Brand\~{a}o}}, \ and\ \bibinfo {author} {\bibfnamefont {K.~M.~R.}\
  \bibnamefont {Audenaert}},\ }\href@noop {} {\bibfield  {journal} {\bibinfo
  {journal} {New J. Phys.}\ }\textbf {\bibinfo {volume} {9}},\ \bibinfo {pages}
  {46} (\bibinfo {year} {2007})}\BibitemShut {NoStop}%
\bibitem [{\citenamefont {Chru\'{s}ci\'{n}ski}\ and\ \citenamefont
  {Sarbicki}(2014)}]{Review}%
  \BibitemOpen
  \bibfield  {author} {\bibinfo {author} {\bibfnamefont {D.}~\bibnamefont
  {Chru\'{s}ci\'{n}ski}}\ and\ \bibinfo {author} {\bibfnamefont
  {G.}~\bibnamefont {Sarbicki}},\ }\href@noop {} {\bibfield  {journal}
  {\bibinfo  {journal} {J. Phys. A: Math. Theor.}\ }\textbf {\bibinfo {volume}
  {47}} (\bibinfo {year} {2014})}\BibitemShut {NoStop}%
\bibitem [{\citenamefont {Peres}(1996)}]{PeresPH}%
  \BibitemOpen
  \bibfield  {author} {\bibinfo {author} {\bibfnamefont {A.}~\bibnamefont
  {Peres}},\ }\href@noop {} {\bibfield  {journal} {\bibinfo  {journal} {Phys.
  Rev. Lett.}\ }\textbf {\bibinfo {volume} {77}},\ \bibinfo {pages} {1413}
  (\bibinfo {year} {1996})}\BibitemShut {NoStop}%
\bibitem [{\citenamefont {Horodecki}\ \emph {et~al.}(1996)\citenamefont
  {Horodecki}, \citenamefont {Horodecki},\ and\ \citenamefont
  {Horodecki}}]{HorodeckiPH}%
  \BibitemOpen
  \bibfield  {author} {\bibinfo {author} {\bibfnamefont {M.}~\bibnamefont
  {Horodecki}}, \bibinfo {author} {\bibfnamefont {P.}~\bibnamefont
  {Horodecki}}, \ and\ \bibinfo {author} {\bibfnamefont {R.}~\bibnamefont
  {Horodecki}},\ }\href@noop {} {\bibfield  {journal} {\bibinfo  {journal}
  {Phys. Lett. A}\ }\textbf {\bibinfo {volume} {223}},\ \bibinfo {pages} {1}
  (\bibinfo {year} {1996})}\BibitemShut {NoStop}%
\bibitem [{\citenamefont {Curty}\ \emph {et~al.}(2004)\citenamefont {Curty},
  \citenamefont {Lewenstein},\ and\ \citenamefont {L\"utkenhaus}}]{EW4}%
  \BibitemOpen
  \bibfield  {author} {\bibinfo {author} {\bibfnamefont {M.}~\bibnamefont
  {Curty}}, \bibinfo {author} {\bibfnamefont {M.}~\bibnamefont {Lewenstein}}, \
  and\ \bibinfo {author} {\bibfnamefont {N.}~\bibnamefont {L\"utkenhaus}},\
  }\href@noop {} {\bibfield  {journal} {\bibinfo  {journal} {Phys. Rev. Lett.}\
  }\textbf {\bibinfo {volume} {92}},\ \bibinfo {pages} {217903} (\bibinfo
  {year} {2004})}\BibitemShut {NoStop}%
\bibitem [{\citenamefont {Curty}\ \emph {et~al.}(2005)\citenamefont {Curty},
  \citenamefont {G\"uhne}, \citenamefont {Lewenstein},\ and\ \citenamefont
  {L\"utkenhaus}}]{EW4_2}%
  \BibitemOpen
  \bibfield  {author} {\bibinfo {author} {\bibfnamefont {M.}~\bibnamefont
  {Curty}}, \bibinfo {author} {\bibfnamefont {O.}~\bibnamefont {G\"uhne}},
  \bibinfo {author} {\bibfnamefont {M.}~\bibnamefont {Lewenstein}}, \ and\
  \bibinfo {author} {\bibfnamefont {N.}~\bibnamefont {L\"utkenhaus}},\
  }\href@noop {} {\bibfield  {journal} {\bibinfo  {journal} {Phys. Rev. A}\
  }\textbf {\bibinfo {volume} {71}},\ \bibinfo {pages} {022306} (\bibinfo
  {year} {2005})}\BibitemShut {NoStop}%
\bibitem [{\citenamefont {Hyllus}\ \emph {et~al.}(2005)\citenamefont {Hyllus},
  \citenamefont {G\"uhne}, \citenamefont {Bru\ss{}},\ and\ \citenamefont
  {Lewenstein}}]{Hyllus2005}%
  \BibitemOpen
  \bibfield  {author} {\bibinfo {author} {\bibfnamefont {P.}~\bibnamefont
  {Hyllus}}, \bibinfo {author} {\bibfnamefont {O.}~\bibnamefont {G\"uhne}},
  \bibinfo {author} {\bibfnamefont {D.}~\bibnamefont {Bru\ss{}}}, \ and\
  \bibinfo {author} {\bibfnamefont {M.}~\bibnamefont {Lewenstein}},\
  }\href@noop {} {\bibfield  {journal} {\bibinfo  {journal} {Phys. Rev. A}\
  }\textbf {\bibinfo {volume} {72}},\ \bibinfo {pages} {012321} (\bibinfo
  {year} {2005})}\BibitemShut {NoStop}%
\bibitem [{\citenamefont {Park}\ \emph {et~al.}(2010)\citenamefont {Park},
  \citenamefont {Lee}, \citenamefont {Kim}, \citenamefont {Choi},\ and\
  \citenamefont {Sim}}]{Park2010}%
  \BibitemOpen
  \bibfield  {author} {\bibinfo {author} {\bibfnamefont {H.~S.}\ \bibnamefont
  {Park}}, \bibinfo {author} {\bibfnamefont {S.-S.~B.}\ \bibnamefont {Lee}},
  \bibinfo {author} {\bibfnamefont {H.}~\bibnamefont {Kim}}, \bibinfo {author}
  {\bibfnamefont {S.-K.}\ \bibnamefont {Choi}}, \ and\ \bibinfo {author}
  {\bibfnamefont {H.-S.}\ \bibnamefont {Sim}},\ }\href@noop {} {\bibfield
  {journal} {\bibinfo  {journal} {Phys. Rev. Lett.}\ }\textbf {\bibinfo
  {volume} {105}},\ \bibinfo {pages} {230404} (\bibinfo {year}
  {2010})}\BibitemShut {NoStop}%
\bibitem [{\citenamefont {Shi}\ \emph {et~al.}(2011)\citenamefont {Shi},
  \citenamefont {Yang}, \citenamefont {Jiang},\ and\ \citenamefont
  {Du}}]{Shi2011}%
  \BibitemOpen
  \bibfield  {author} {\bibinfo {author} {\bibfnamefont {M.}~\bibnamefont
  {Shi}}, \bibinfo {author} {\bibfnamefont {W.}~\bibnamefont {Yang}}, \bibinfo
  {author} {\bibfnamefont {F.}~\bibnamefont {Jiang}}, \ and\ \bibinfo {author}
  {\bibfnamefont {J.}~\bibnamefont {Du}},\ }\href@noop {} {\bibfield  {journal}
  {\bibinfo  {journal} {J.\ Phys.\ A: Math.\ Theor.}\ }\textbf {\bibinfo
  {volume} {44}},\ \bibinfo {pages} {415304} (\bibinfo {year}
  {2011})}\BibitemShut {NoStop}%
\bibitem [{\citenamefont {Verstraete}(2002)}]{Verstraete2002}%
  \BibitemOpen
  \bibfield  {author} {\bibinfo {author} {\bibfnamefont {F.}~\bibnamefont
  {Verstraete}},\ }\href@noop {} {Ph.D. thesis},\ \bibinfo  {school}
  {Katholieke Universiteit Leuven} (\bibinfo {year} {2002})\BibitemShut
  {NoStop}%
\bibitem [{\citenamefont {Jevtic}\ \emph {et~al.}(2014)\citenamefont {Jevtic},
  \citenamefont {Pusey}, \citenamefont {Jennings},\ and\ \citenamefont
  {Rudolph}}]{Jevtic2013}%
  \BibitemOpen
  \bibfield  {author} {\bibinfo {author} {\bibfnamefont {S.}~\bibnamefont
  {Jevtic}}, \bibinfo {author} {\bibfnamefont {M.}~\bibnamefont {Pusey}},
  \bibinfo {author} {\bibfnamefont {D.}~\bibnamefont {Jennings}}, \ and\
  \bibinfo {author} {\bibfnamefont {T.}~\bibnamefont {Rudolph}},\ }\href@noop
  {} {\bibfield  {journal} {\bibinfo  {journal} {Phys. Rev. Lett.}\ }\textbf
  {\bibinfo {volume} {113}},\ \bibinfo {pages} {020402} (\bibinfo {year}
  {2014})}\BibitemShut {NoStop}%
\bibitem [{\citenamefont {Altepeter}\ \emph {et~al.}(2009)\citenamefont
  {Altepeter}, \citenamefont {Jeffrey}, \citenamefont {Medic},\ and\
  \citenamefont {Kumar}}]{Altepeter09}%
  \BibitemOpen
  \bibfield  {author} {\bibinfo {author} {\bibfnamefont {J.}~\bibnamefont
  {Altepeter}}, \bibinfo {author} {\bibfnamefont {E.}~\bibnamefont {Jeffrey}},
  \bibinfo {author} {\bibfnamefont {M.}~\bibnamefont {Medic}}, \ and\ \bibinfo
  {author} {\bibfnamefont {P.}~\bibnamefont {Kumar}},\ }in\ \href@noop {}
  {\emph {\bibinfo {booktitle} {Conference on Lasers and
  Electro-Optics/International Quantum Electronics Conference}}}\ (\bibinfo
  {publisher} {Optical Society of America},\ \bibinfo {year} {2009})\ p.\
  \bibinfo {pages} {IWC1}\BibitemShut {NoStop}%
\bibitem [{\citenamefont {Milne}\ \emph
  {et~al.}(2014{\natexlab{a}})\citenamefont {Milne}, \citenamefont {Jevtic},
  \citenamefont {Jennings}, \citenamefont {Wiseman},\ and\ \citenamefont
  {Rudolph}}]{Monogamy}%
  \BibitemOpen
  \bibfield  {author} {\bibinfo {author} {\bibfnamefont {A.}~\bibnamefont
  {Milne}}, \bibinfo {author} {\bibfnamefont {S.}~\bibnamefont {Jevtic}},
  \bibinfo {author} {\bibfnamefont {D.}~\bibnamefont {Jennings}}, \bibinfo
  {author} {\bibfnamefont {H.}~\bibnamefont {Wiseman}}, \ and\ \bibinfo
  {author} {\bibfnamefont {T.}~\bibnamefont {Rudolph}},\ }\href@noop {}
  {\bibfield  {journal} {\bibinfo  {journal} {New J. Phys.}\ }\textbf {\bibinfo
  {volume} {16}},\ \bibinfo {pages} {083017} (\bibinfo {year}
  {2014}{\natexlab{a}})}\BibitemShut {NoStop}%
\bibitem [{\citenamefont {Milne}\ \emph
  {et~al.}(2014{\natexlab{b}})\citenamefont {Milne}, \citenamefont {Jennings},
  \citenamefont {Jevtic},\ and\ \citenamefont {Rudolph}}]{Obesity}%
  \BibitemOpen
  \bibfield  {author} {\bibinfo {author} {\bibfnamefont {A.}~\bibnamefont
  {Milne}}, \bibinfo {author} {\bibfnamefont {D.}~\bibnamefont {Jennings}},
  \bibinfo {author} {\bibfnamefont {S.}~\bibnamefont {Jevtic}}, \ and\ \bibinfo
  {author} {\bibfnamefont {T.}~\bibnamefont {Rudolph}},\ }\href@noop {}
  {\bibfield  {journal} {\bibinfo  {journal} {Phys. Rev. A}\ }\textbf {\bibinfo
  {volume} {90}},\ \bibinfo {pages} {024302} (\bibinfo {year}
  {2014}{\natexlab{b}})}\BibitemShut {NoStop}%
\bibitem [{\citenamefont {Terhal}(2001)}]{Terhal2000}%
  \BibitemOpen
  \bibfield  {author} {\bibinfo {author} {\bibfnamefont {B.~M.}\ \bibnamefont
  {Terhal}},\ }\href@noop {} {\bibfield  {journal} {\bibinfo  {journal} {Linear
  Alg. Appl.}\ }\textbf {\bibinfo {volume} {323}},\ \bibinfo {pages} {61}
  (\bibinfo {year} {2001})}\BibitemShut {NoStop}%
\bibitem [{\citenamefont {Lewenstein}\ \emph {et~al.}(2000)\citenamefont
  {Lewenstein}, \citenamefont {Kraus}, \citenamefont {Cirac},\ and\
  \citenamefont {Horodecki}}]{OptimizationEW}%
  \BibitemOpen
  \bibfield  {author} {\bibinfo {author} {\bibfnamefont {M.}~\bibnamefont
  {Lewenstein}}, \bibinfo {author} {\bibfnamefont {B.}~\bibnamefont {Kraus}},
  \bibinfo {author} {\bibfnamefont {J.~I.}\ \bibnamefont {Cirac}}, \ and\
  \bibinfo {author} {\bibfnamefont {P.}~\bibnamefont {Horodecki}},\ }\href@noop
  {} {\bibfield  {journal} {\bibinfo  {journal} {Phys. Rev. A}\ }\textbf
  {\bibinfo {volume} {62}},\ \bibinfo {pages} {052310} (\bibinfo {year}
  {2000})}\BibitemShut {NoStop}%
\bibitem [{Note1()}]{Note1}%
  \BibitemOpen
  \bibinfo {note} {Slightly abusing notation, we use the term `Bloch sphere' to
  mean the unit ball with centre $\protect \bm {0}$. Thus by `inside' we
  include the possibility that $\protect \cal E$ is the whole Bloch
  sphere.}\BibitemShut {Stop}%
\bibitem [{Note2()}]{Note2}%
  \BibitemOpen
  \bibinfo {note} {When $R_B$ is singular ($b=1$) the canonical transformation
  cannot be performed; \protect \textup {\hbox {\mathsurround \z@ \protect
  \normalfont (\ignorespaces \ref {eq:centre}\unskip \@@italiccorr )}} and
  \protect \textup {\hbox {\mathsurround \z@ \protect \normalfont
  (\ignorespaces \ref {eq:Qmatrix}\unskip \@@italiccorr )}} do not then apply,
  and $\protect \cal E$ is simply a point at $\protect \bm {a}$. This
  corresponds to $R$ being a product state (given that $\protect \cal E$ lies
  within the Bloch sphere).}\BibitemShut {Stop}%
\bibitem [{\citenamefont {Avron}\ \emph {et~al.}(2007)\citenamefont {Avron},
  \citenamefont {Bisker},\ and\ \citenamefont {Kenneth}}]{Avron2007}%
  \BibitemOpen
  \bibfield  {author} {\bibinfo {author} {\bibfnamefont {J.~E.}\ \bibnamefont
  {Avron}}, \bibinfo {author} {\bibfnamefont {G.}~\bibnamefont {Bisker}}, \
  and\ \bibinfo {author} {\bibfnamefont {O.}~\bibnamefont {Kenneth}},\
  }\href@noop {} {\bibfield  {journal} {\bibinfo  {journal} {J. Math. Phys.}\
  }\textbf {\bibinfo {volume} {48}},\ \bibinfo {pages} {102107} (\bibinfo
  {year} {2007})}\BibitemShut {NoStop}%
\bibitem [{\citenamefont {Bryant}()}]{MathOverflow}%
  \BibitemOpen
  \bibfield  {author} {\bibinfo {author} {\bibfnamefont {R.}~\bibnamefont
  {Bryant}},\ }\href@noop {} {\enquote {\bibinfo {title} {Conditions for when
  an off-centre ellipsoid fits inside the unit ball},}\ }\bibinfo
  {howpublished} {\url{http://mathoverflow.net/questions/140339/}},\ \bibinfo
  {note} {accessed 5 June 2014}\BibitemShut {NoStop}%
\bibitem [{\citenamefont {Johnston}(2012)}]{BlockP}%
  \BibitemOpen
  \bibfield  {author} {\bibinfo {author} {\bibfnamefont {N.}~\bibnamefont
  {Johnston}},\ }\href@noop {} {Ph.D. thesis},\ \bibinfo  {school} {University
  of Guelph} (\bibinfo {year} {2012})\BibitemShut {NoStop}%
\bibitem [{\citenamefont {Woronowicz}(1976)}]{Pmap1}%
  \BibitemOpen
  \bibfield  {author} {\bibinfo {author} {\bibfnamefont {S.~L.}\ \bibnamefont
  {Woronowicz}},\ }\href@noop {} {\bibfield  {journal} {\bibinfo  {journal}
  {Rep. Math. Phys.}\ }\textbf {\bibinfo {volume} {10}},\ \bibinfo {pages}
  {165} (\bibinfo {year} {1976})}\BibitemShut {NoStop}%
\bibitem [{\citenamefont {\.{Z}yczkowski}\ and\ \citenamefont
  {Bengtsson}(2004)}]{Pmap2}%
  \BibitemOpen
  \bibfield  {author} {\bibinfo {author} {\bibfnamefont {K.}~\bibnamefont
  {\.{Z}yczkowski}}\ and\ \bibinfo {author} {\bibfnamefont {I.}~\bibnamefont
  {Bengtsson}},\ }\href@noop {} {\bibfield  {journal} {\bibinfo  {journal}
  {Open Syst. Inf. Dyn.}\ }\textbf {\bibinfo {volume} {11}},\ \bibinfo {pages}
  {3} (\bibinfo {year} {2004})}\BibitemShut {NoStop}%
\bibitem [{\citenamefont {Sarbicki}(2008)}]{Sarbicki2008}%
  \BibitemOpen
  \bibfield  {author} {\bibinfo {author} {\bibfnamefont {G.}~\bibnamefont
  {Sarbicki}},\ }\href@noop {} {\bibfield  {journal} {\bibinfo  {journal} {J.
  Phys. A: Math. Theor.}\ }\textbf {\bibinfo {volume} {41}},\ \bibinfo {pages}
  {375303} (\bibinfo {year} {2008})}\BibitemShut {NoStop}%
\bibitem [{\citenamefont {Augusiak}\ \emph {et~al.}(2008)\citenamefont
  {Augusiak}, \citenamefont {Demianowicz},\ and\ \citenamefont
  {Horodecki}}]{detrhoref}%
  \BibitemOpen
  \bibfield  {author} {\bibinfo {author} {\bibfnamefont {R.}~\bibnamefont
  {Augusiak}}, \bibinfo {author} {\bibfnamefont {M.}~\bibnamefont
  {Demianowicz}}, \ and\ \bibinfo {author} {\bibfnamefont {P.}~\bibnamefont
  {Horodecki}},\ }\href@noop {} {\bibfield  {journal} {\bibinfo  {journal}
  {Phys. Rev. A}\ }\textbf {\bibinfo {volume} {77}},\ \bibinfo {pages} {030301}
  (\bibinfo {year} {2008})}\BibitemShut {NoStop}%
\bibitem [{\citenamefont {Braun}\ \emph {et~al.}(2014)\citenamefont {Braun},
  \citenamefont {Giraud}, \citenamefont {Nechita}, \citenamefont {Pellegrini},\
  and\ \citenamefont {Znidaric}}]{Braun2013}%
  \BibitemOpen
  \bibfield  {author} {\bibinfo {author} {\bibfnamefont {D.}~\bibnamefont
  {Braun}}, \bibinfo {author} {\bibfnamefont {O.}~\bibnamefont {Giraud}},
  \bibinfo {author} {\bibfnamefont {I.}~\bibnamefont {Nechita}}, \bibinfo
  {author} {\bibfnamefont {C.}~\bibnamefont {Pellegrini}}, \ and\ \bibinfo
  {author} {\bibfnamefont {M.}~\bibnamefont {Znidaric}},\ }\href@noop {}
  {\bibfield  {journal} {\bibinfo  {journal} {J. Phys. A: Math. Theor.}\
  }\textbf {\bibinfo {volume} {47}},\ \bibinfo {pages} {135302} (\bibinfo
  {year} {2014})}\BibitemShut {NoStop}%
\bibitem [{\citenamefont {Badziag}\ \emph {et~al.}(2013)\citenamefont
  {Badziag}, \citenamefont {Horodecki}, \citenamefont {Horodecki},\ and\
  \citenamefont {Augusiak}}]{WeaklyOptimal}%
  \BibitemOpen
  \bibfield  {author} {\bibinfo {author} {\bibfnamefont {P.}~\bibnamefont
  {Badziag}}, \bibinfo {author} {\bibfnamefont {P.}~\bibnamefont {Horodecki}},
  \bibinfo {author} {\bibfnamefont {R.}~\bibnamefont {Horodecki}}, \ and\
  \bibinfo {author} {\bibfnamefont {R.}~\bibnamefont {Augusiak}},\ }\href@noop
  {} {\bibfield  {journal} {\bibinfo  {journal} {Phys. Rev. A}\ }\textbf
  {\bibinfo {volume} {88}},\ \bibinfo {pages} {010301} (\bibinfo {year}
  {2013})}\BibitemShut {NoStop}%
\bibitem [{\citenamefont {Korbicz}\ \emph {et~al.}(2008)\citenamefont
  {Korbicz}, \citenamefont {Almeida}, \citenamefont {Bae}, \citenamefont
  {Lewenstein},\ and\ \citenamefont {Ac\'in}}]{Korbicz2008}%
  \BibitemOpen
  \bibfield  {author} {\bibinfo {author} {\bibfnamefont {J.~K.}\ \bibnamefont
  {Korbicz}}, \bibinfo {author} {\bibfnamefont {M.~L.}\ \bibnamefont
  {Almeida}}, \bibinfo {author} {\bibfnamefont {J.}~\bibnamefont {Bae}},
  \bibinfo {author} {\bibfnamefont {M.}~\bibnamefont {Lewenstein}}, \ and\
  \bibinfo {author} {\bibfnamefont {A.}~\bibnamefont {Ac\'in}},\ }\href@noop {}
  {\bibfield  {journal} {\bibinfo  {journal} {Phys. Rev. A}\ }\textbf {\bibinfo
  {volume} {78}},\ \bibinfo {pages} {062105} (\bibinfo {year}
  {2008})}\BibitemShut {NoStop}%
\bibitem [{\citenamefont {Wang}\ \emph {et~al.}(2014)\citenamefont {Wang},
  \citenamefont {Xu}, \citenamefont {Campbell},\ and\ \citenamefont
  {Severini}}]{Wang2014}%
  \BibitemOpen
  \bibfield  {author} {\bibinfo {author} {\bibfnamefont {B.-H.}\ \bibnamefont
  {Wang}}, \bibinfo {author} {\bibfnamefont {H.-R.}\ \bibnamefont {Xu}},
  \bibinfo {author} {\bibfnamefont {S.}~\bibnamefont {Campbell}}, \ and\
  \bibinfo {author} {\bibfnamefont {S.}~\bibnamefont {Severini}},\ }\href@noop
  {} {\enquote {\bibinfo {title} {Characterization and properties of weakly
  optimal entanglement witnesses},}\ } (\bibinfo {year} {2014}),\ \Eprint
  {http://arxiv.org/abs/quant-ph/1407.0870} {quant-ph/1407.0870} \BibitemShut
  {NoStop}%
\bibitem [{\citenamefont {Wang}\ and\ \citenamefont {Long}(2013)}]{Wang2013}%
  \BibitemOpen
  \bibfield  {author} {\bibinfo {author} {\bibfnamefont {B.-H.}\ \bibnamefont
  {Wang}}\ and\ \bibinfo {author} {\bibfnamefont {D.-Y.}\ \bibnamefont
  {Long}},\ }\href@noop {} {\bibfield  {journal} {\bibinfo  {journal} {Phys.
  Rev. A}\ }\textbf {\bibinfo {volume} {87}},\ \bibinfo {pages} {062324}
  (\bibinfo {year} {2013})}\BibitemShut {NoStop}%
\bibitem [{\citenamefont {Ried}\ \emph {et~al.}(2015)\citenamefont {Ried},
  \citenamefont {Agnew}, \citenamefont {Vermeyden}, \citenamefont {Janzing},
  \citenamefont {Spekkens},\ and\ \citenamefont {Resch}}]{Ried2014}%
  \BibitemOpen
  \bibfield  {author} {\bibinfo {author} {\bibfnamefont {K.}~\bibnamefont
  {Ried}}, \bibinfo {author} {\bibfnamefont {M.}~\bibnamefont {Agnew}},
  \bibinfo {author} {\bibfnamefont {L.}~\bibnamefont {Vermeyden}}, \bibinfo
  {author} {\bibfnamefont {D.}~\bibnamefont {Janzing}}, \bibinfo {author}
  {\bibfnamefont {R.~W.}\ \bibnamefont {Spekkens}}, \ and\ \bibinfo {author}
  {\bibfnamefont {K.~J.}\ \bibnamefont {Resch}},\ }\href@noop {} {\bibfield
  {journal} {\bibinfo  {journal} {Nat. Phys.}\ }\textbf {\bibinfo {volume}
  {11}},\ \bibinfo {pages} {414} (\bibinfo {year} {2015})}\BibitemShut
  {NoStop}%
\end{thebibliography}%

\end{document}